\documentstyle[12pt]{article}

\newcommand\Journal[4]{{#1} {\bf #2}, #3 (#4)}
\newcommand\NCA{\em Nuovo Cimento}

\newcommand\NPB{{\em Nucl. Phys.} B}
\newcommand\PLB{{\em Phys. Lett.}  B}
\newcommand\PRL{\em Phys. Rev. Lett.}
\newcommand\PRD{{\em Phys. Rev.} D}
\newcommand\ZPC{{\em Z. Phys.} C}
\newcommand\PR{{\em Phys. Rep.}}
\newcommand\IMPA{{\em Int. J. Mod. Phys.} A}
\newcommand\ARNPS{{\em Annu. Rev. Nucl. Part. Sci.} }
\newcommand\NPPS{{\em Nucl. Phys. Proc. Suppl.} }

\newcounter{two}
\def\hybrid{\topmargin +10pt    \oddsidemargin 0pt
\headheight 0pt \headsep 0pt
       \textwidth 6.5in        
       \textheight 9in         
        \marginparwidth .875in \parskip 5pt plus 1pt   \jot = 1.5ex}
        \hybrid
        \newcommand{\PSbox}[3]{\mbox{\rule{0in}{#3}\includegraphics{#1}\hspace{#2}}}

\begin{document}

\begin{titlepage}
\begin{flushright}
MIT-CTP-2755\\
CERN-TH/98-159\\
hep-ph/yymmddd
\end{flushright}
\LARGE
\vspace{0.3in}

\begin{center}
{\bf  CP Violating Lepton Asymmetries from $B$ Decays and Their Implication
for  Supersymmetric Flavor Models} \\
\vspace{0.4in}

\normalsize

{Lisa Randall and Shufang Su}\\
\vspace{.05in}
{ \small \it Center for Theoretical Physics\\
Laboratory for Nuclear Science and Department of Physics\\
Massachusetts Institute of Technology\\
Cambridge, MA 02139, USA }
\end{center}
 
\vspace{0.2in}
 
\vspace{0.25in}
\LARGE
\begin{center}
Abstract
\end{center}
\normalsize   
The lepton  and dilepton charge asymmetries from   $B_{d}$
and $B_{s}$ are predicted to be   small in the
standard model, whereas  new physics could increase their
values significantly. In this paper, we explore the
use of the lepton asymmetries as a probe of
the flavor structure of supersymmetric theories.
In particular, we 
determine the sensitivity to parameters of various models.  
We find that in many interesting models which attempt
to address the supersymmetric flavor problem, the mixing
structure is such that it could be possible
to detect new physics. The predictions are model
dependent; with a measurement in both the $B_s$ and $B_d$
systems one can hope to constrain the flavor physics
model, especially once squarks are detected and their masses measured.         
Thus,  lepton  charge asymmetries
can be used as an alternative means of searching for new physics and
distinguishing among potential
solutions   to the flavor problem. They
are interesting precisely because they are small
in the standard model and are therefore   necessarily   evidence
of new physics.

\end{titlepage}

\section{Introduction}
\label{introduction}
\setcounter{equation}{0}
The next few years will be an exciting era for $B$ physics,
with the detailed investigation of $B$ hadrons at  
$b$-factories. Particularly exciting is the potential for
studying CP-violation  in the $B$ system, both within and
beyond the standard model (SM). This affords the opportunity
to look for new physics, and in fact might yield
the first evidence for physics beyond the standard model.
Hopefully this new physics will be further studied
directly so that we will establish its origin.
Whatever this new physics proves to be, a detection
at $b$-factories should give new information which
will not be accessible to high energy colliders.
For example, should this new physics prove to be supersymmetry,
detailed studies in the $B$ system give   a unique opportunity
to probe the flavor structure of extensions of the SM.

Non-standard model CP-violating
effects  could  be revealed by testing whether measurements
agree with the SM allowed range. 
Processes for which the SM
contribution vanishes (or is negligibly small)  offer
an important complement to these studies. In  this case, any observation
or non-observation of CP-violation can be interpreted
directly as a constraint on physics beyond the SM.
From this point of view, a measurement of the dilepton or of the lepton 
charge asymmetry 
is of great interest. The dilepton charge asymmetry is defined 
as\footnote{Here we use the convention that $B$ contains a $\bar{b}$ quark, 
thus decays into a $l^+$ if there is no mixing.  This is opposite to the 
convention used in \cite{franzini}.}
\cite{okun,franzini,hag,acuto,bss,yamamoto}
\begin{equation}
A_{ll}\equiv\frac{l^{++}-l^{--}}{l^{++}+l^{--}},
\label{didef}
\end{equation}
where $l^{++}$ [$l^{--}$] denotes the numbers of $l^{+}l^{+}$ 
[$l^{-}l^{-}$]  dilepton
pairs observed.  In the $b$-factories, they come from
the  mixing and decay of the $B\bar{B}$ pairs,
while in the hadron colliders, the final
hadronization states can be any combinations of $B^+,\ B_{d},\ B_{s},
\ \Lambda_{b}$ and their conjugates.  In the absence of
CP-violation, this quantity clearly vanishes.  In early studies of the
dilepton charge asymmetry \cite{bigi,nirslac},  the KM angles and top
quark mass were not sufficiently well determined to be  certain that a
measurement in excess of $10^{-3}$ would signal new physics.  As we
will  see, this quantity is now determined  to be small in the SM for
both $B_d$ and $B_s$, but can be significantly larger in non-standard
flavor models.

Another useful quantity to look at \cite{franzini,hag,bss,yamamoto} is the
total lepton charge asymmetry $l^{\pm}$, which is defined by
\begin{equation}
l^{\pm} ={l^+-l^- \over l^++l^-}.
\end{equation}
Here $l^{+(-)}=N(B\bar{B}\rightarrow{l}^{+(-)})$ 
is the total number of positively (negatively) charged
primary leptons coming from the decay of $b\bar{b}$ pairs.
This quantity is smaller than $A_{ll}$, but should be measured
with better statistics.

These lepton asymmetries are sensitive to the phase difference between
$\Gamma_{12}$
and $M_{12}$. In the SM, the dominant contribution  to
these quantities has the same 
phase, and is therefore   
suppressed.  The dominant
source of enhancement in non-standard physics is 
a new contribution to
mass mixing, which  would generally carry a different phase
from the standard model contribution.    In models
for which the new source of mixing is comparable to that of the
SM, one can expect a substantial change in the prediction
of the lepton charge asymmetry.
We will show that there
are many models for which one would expect up to  an order
of magnitude enhancement over the SM prediction for $B_d$
and two orders of magnitude enhancement for $B_s$. 
In fact, because the predictions for asymmetries due to CP
violation in the $B_s$ and $B_d$ systems is different,
one can hope to use this measurement to help distinguish
among potential solutions to the supersymmetric flavor problem.
For this reason, we take an explicitly model-dependent
approach to our results, and explore the predictions
in various models already existing in the literature.
They do not necessarily include the real world solutions to the flavor 
problems, but are nonetheless sufficiently general 
to illustrate the usefulness of 
the lepton asymmetry methods.

There are several ways to measure the lepton asymmetries.
One can measure both the single lepton and dilepton asymmetries
at the dedicated $B$ factories. These will of course only
be sensitive to new physics in the $B_d$ sector. It would
be extremely interesting to complement this measurement with
the measurement of lepton asymmetries at a hadron collider,
which will be sensitive to the asymmetries in both $B_s$ and $B_d$. 
With all such
measurements (or even some fraction of them) one should be 
able to distinguish new mixing  contributions to either the
$B_s$ or $B_d$ systems. We will see in particular that many supersymmetric
flavor models yield a large deviation for at least one of the above
systems. 
 
Because
the SM prediction for $B_d$ is small, and even smaller for
$B_s$ (we discuss how small later), any lepton
asymmetry measurement in excess of this value is a clear
signal of new physics. Because of the suppression from $\Delta \Gamma/
\Delta M$, a sensitivity of at least  $10^{-2}$ is essential. So
any asymmetry within this range will be an important signal of new physics.
We emphasize  that even
without flavor tagging, a measurement of a CP-violating asymmetry
in excess of the SM prediction 
will be an exciting signal of new physics, which ultimately complementary
measurements should disentangle.

The importance of studying lepton asymmetries has been considered
previously  in \cite{xing2} specifically in the context of  $B_s$, and
in \cite{xing1} for $B_d$.  ${\rm Re}(\bar{\epsilon}_{B})$ (which can
be related to $A_{ll}$ by $A_{ll}\sim{4}{\rm Re}(\bar{\epsilon}_{B})$ 
\cite{bernabeu1}) in left-right symmetric models has also been
discussed in \cite{bernabeu}. 
However, no one has
as yet done  a detailed study   of the potential
significance of this measurement for distinguishing
models for the supersymmetric soft masses or combined 
the information from both $B_d$ and $B_s$.  Of course,
any non-standard model can be studied in the light of the measurement
of the lepton charge asymmetry and thereby constrained.   
For specificity, and because of its likelihood as the source
of a non-standard model CP-violating effect, we chose to study the 
specific case of contributions
from soft scalar masses in some non-standard models
of flavor physics that exist in the literature.  
We find that many squark mass models designed to
address the flavor problem in supersymmetry  will give rise to a 
significantly larger lepton asymmetry in either the $B_d$ or $B_s$
system  for reasonable parameters. A simple order of magnitude
estimate shows that the box diagrams would be comparable
(assuming mixing similar to CKM mixing) for superpartner
masses of order a few hundred GeV.

We begin in Sec.~\ref{formula}  by presenting  the basic formulas  relevant to 
$B\bar{B}$ mixing.  We express the dilepton asymmetry in terms of  
$(\Delta\Gamma/\Delta{M})_{\rm SM}$, and the relative 
amplitude and phase difference  between the supersymmetric 
and SM contributions.  We then
briefly review why supersymmetry gives new contributions
to neutral $B$-meson mixing through non-trivial squark mass
matrices. In Sec.~\ref{model},
we give a brief review of three soft supersymmetry-breaking scenarios
which have been devised to address the flavor problem in supersymmetry models:
alignment, non-abelian  flavor symmetry,
and heavy squark models. These suggestions  
solve or relax the Flavor-Changing Neutral-Current (FCNC)
and CP-violation problems.  In
Sec.~\ref{numerical}, we first present the model-independent 
lower limits on the phase
difference between $M_{12}^{\rm SUSY}$ and $M_{12}^{\rm SM}$  and the 
$m_{\tilde{q}}$ range (linear
with respect to the mixing angle), assuming the experimental
sensitivity of measuring either dilepton asymmetry is $2\times{10}^{-3}$.  
For any particular ansatz for the squark mass matrices,
this result can be interpreted as a sensitivity
to  $m_{\tilde{q}}$ and   mixing angles. We show
the parameter ranges  of the models of Sec. \ref{model}, which
can be probed with the measurement of the dilepton asymmetry.    
In Sec.~\ref{conclusion}, we conclude.

\section{Dilepton Charge Asymmetry}
\label{formula}
\setcounter{equation}{0} For the $B$ $\bar{B}$ basis, one has the
Hamiltonian
\begin{equation}
H=\left(
\begin{array}{cc}
M-{\rm i}\Gamma/2&M_{12}-{\rm i}\Gamma_{12}/2\\ M_{12}^*-{\rm
i}\Gamma_{12}^*/2&M-{\rm i}\Gamma/2 \,.
\end{array}\right)
\end{equation}
The mass eigenstates are
\begin{equation}
B_{1,2}=\frac{1}{\sqrt{1+|\eta|^2}}(|B\rangle\pm\eta|\bar{B}\rangle),
\end{equation}
with eigenvalues
\begin{equation}
M_{1,2}-{\rm i}\Gamma_{1,2}/2=M-{\rm i}\Gamma/2\pm\Delta,
\end{equation}
where
\begin{equation}
\eta=\sqrt{\frac{M_{12}^*-{\rm i}\Gamma_{12}^*/2}{M_{12}-{\rm i}
\Gamma_{12}/2}},
\end{equation}
\begin{equation}
\Delta=\sqrt{(M_{12}-{\rm i}\Gamma_{12}/2)(M_{12}^*-{\rm
i}\Gamma_{12}^*/2)}.
\end{equation}
The quantities $r$,$\bar{r}$  are defined as \cite{franzini,pais}
\footnote{The $r$, $\bar{r}$ defined here is the interchange of  $r$,
$\bar{r}$ defined in \cite{franzini,pais} because of the opposite
convention used in defining $B$.}
\begin{equation}
r\equiv\frac{P_{\bar{B}\rightarrow{B}}}{P_{\bar{B}\rightarrow\bar{B}}}
=\frac{1}{|\eta|^2}\frac{x^2+y^2}{2+x^2-y^2},
\label{defr1}
\end{equation}
\begin{equation}
\bar{r}\equiv\frac{P_{B\rightarrow\bar{B}}}{P_{B\rightarrow{B}}}
=|\eta|^2\frac{x^2+y^2}{2+x^2-y^2},
\label{defr2}
\end{equation}
where $x=\Delta{M}/\Gamma$, $y=\Delta\Gamma/2\Gamma$,
$\Delta{M}=M_1-M_2 =2{\rm Re}\Delta$, $\Gamma=(\Gamma_1+\Gamma_2)/2$
and  $\Delta\Gamma=\Gamma_1-\Gamma_2=-4{\rm Im}\Delta$.

When a $B\bar{B}$ pair is produced, it can mix and later decay into
$l^+l^+$ or $l^-l^-$.  Thus, we can replace the $l^{++}$ and $l^{--}$
in Eq.~(\ref{didef}) by $N(BB)$ and $N(\bar{B}\bar{B})$, which is the 
number of $BB$ ($\bar{B}\bar{B}$) final states observed in a sample of 
events from a process where a $B\bar{B}$ pair is initially produced 
\cite{franzini}.   The
dilepton asymmetry can then be written as \cite{franzini,hag}
\begin{equation}
A_{ll}\equiv{N(BB)-N(\bar{B}\bar{B}) \over N(BB)+N(\bar{B}\bar{B})}
={r-\bar{r} \over r + \bar{r}} =-\frac{|\eta|^4-1}{|\eta|^4+1}
=\frac{{\rm Im}(\Gamma_{12}/M_{12})}{1+1/4|\Gamma_{12}/M_{12}|^2}
\approx{{\rm Im} \left(\frac{\Gamma_{12}}{M_{12}}\right)}.
\end{equation}
The last approximation holds   if $|\Gamma_{12}/M_{12}|\ll1$, which is
the case  for the  $B\bar{B}$ systems even  in  the presence of new
physics \cite{dib}.

This formula is true whether or not $B\bar{B}$ is produced
coherently.  However at a hadron collider when a $B_{d}$ is
not necessarily produced in conjunction with a $\bar{B}_d$,
one needs to account for all possible fragmentations.
In this case we derive
\begin{equation}
A_{ll}=\frac{l^{++}-l^{--}}{l^{++}+l^{--}}=\frac{ \left(\frac{r_d}{1+r_d}
\lambda_d f_d +\frac{r_s}{1+r_s}\lambda_s f_s \right) A^{++}
-\left(\frac{\bar{r}_d}{1+\bar{r}_d} \lambda_d f_d +
\frac{\bar{r}_s}{1+\bar{r}_s} \lambda_s f_s \right) A^{--}}
{\left(\frac{r_d}{1+r_d} \lambda_d f_d +\frac{r_s}{1+r_s}\lambda_s f_s
\right) A^{++} +\left(\frac{\bar{r}_d}{1+\bar{r}_d} \lambda_d f_d +
\frac{\bar{r}_s}{1+\bar{r}_s} \lambda_s f_s \right) A^{--}},
\label{all}
\end{equation}
where
\begin{equation}
A^{++}=\lambda^+ f^++\lambda^\Lambda f^\Lambda+ \frac{\lambda_d
f_d}{1+\bar{r}_d} +\frac{\lambda_s f_s}{1+\bar{r}_s},
\end{equation}
and
\begin{equation}
A^{--}=\lambda^+f^++\lambda^\Lambda f^\Lambda+\frac{\lambda_d
f_d}{1+r_d}+ \frac{\lambda_s f_s}{1+r_s}.
\end{equation}
Here $f^i$ is the probability to hadronize as a state $i$ and
$\lambda^i$ is the leptonic branching fraction.  It is readily seen
that when  working to leading order in CP asymmetries, 
the formula (\ref{all}) reduces to
\begin{equation}
(A_{ll})_{\rm total}=(A_{ll})_{d}A_{d2}+(A_{ll})_{s}A_{s2},
\hspace{0.2 in}{\rm with}\
(A_{ll})_{d,s} =\frac{r_{d,s}-\bar{r}_{d,s}}{r_{d,s}+\bar{r}_{d,s}},
\label{r2total}
\end{equation}
where
\begin{equation}
A_{d2}=\frac{\lambda_df_dr_d}{(1+r_d)^2}\frac{(\lambda^+f^++\lambda^\Lambda
f^\Lambda+\lambda_df_d+\lambda_sf_s)}{(\lambda^+f^++\lambda^\Lambda
f^\Lambda+\frac{\lambda_d f_d}{1+r_d}+ \frac{\lambda_s
f_s}{1+r_s})(\frac{\lambda_d f_dr_d}{1+r_d}+ \frac{\lambda_s
f_sr_s}{1+r_s})},
\end{equation}
\begin{equation}
A_{s2}=\frac{\lambda_sf_sr_s}{(1+r_s)^2}\frac{(\lambda^+f^++\lambda^\Lambda
f^\Lambda+\lambda_df_d+\lambda_sf_s)}{(\lambda^+f^++\lambda^\Lambda
f^\Lambda+\frac{\lambda_d f_d}{1+r_d}+ \frac{\lambda_s
f_s}{1+r_s})(\frac{\lambda_d f_dr_d}{1+r_d}+ \frac{\lambda_s
f_sr_s}{1+r_s})},
\end{equation}
and we define $A_2$ to be the ratio between $A_{s2}$ and $A_{d2}$, which
gives
\begin{equation}
A_2=\frac{A_{s2}}{A_{d2}}=\frac{\lambda_sf_sr_s(1+r_d)^2}
{\lambda_df_dr_d(1+r_s)^2}.
\end{equation}
The most recent results for $f^i$ and $\lambda^i$  are
$f^+=0.39\pm{0.04}\pm{0.04}$, $f^{\Lambda}=0.096\pm{0.017}$,  
$f_{d}=0.38\pm{0.04}\pm{0.04}$, $f_{s}=0.13\pm{0.03}\pm{0.01}$ \cite{cdfb}, 
$\lambda^+=(10.3\pm{0.9})\%$,
$\lambda^{\Lambda}=(9.0^{+3.1}_{-3.8})\%$,
$\lambda_d=(10.5\pm{0.8})\%$ and
$\lambda_s=(8.1\pm{2.5})\%$ \cite{pdb}.  For  the $B\bar{B}$ system,
$|\eta|$ in Eqs.~(\ref{defr1}) and (\ref{defr2}) is very close to 1,
$y/x=\Delta\Gamma/2\Delta{M}$ is very close to 0, thus
\begin{equation}
r\approx\bar{r}\approx\frac{x^2}{2+x^2}.
\end{equation}
For $B_s$, $x$ is large because of large mixing; 
thus $B_{s}$ is almost $100\%$ mixed and $r\rightarrow{1}$.
For $B_d$, $x=0.734\pm0.035$\cite{pdb}, which gives $r=0.21\pm{0.02}$.  
Putting together these numbers, we find
that $A_{d2}=0.53$, $A_{s2}=0.25$, $A_2=0.46$; this means that, 
although $B_s$ is 100\% mixed, its
contribution to the dilepton rate is less than $B_d$'s because
the number of $B_d$s  produced is about  three times larger than $B_s$.  In our
analysis below, we will consider supersymmetric contributions to
both $B_d$ and $B_s$ mixing.  It should be borne in mind that a better 
measurement of $A_{ll}$ is required to achieve the same sensitivity to 
the parameters relevant to $B_s$.

The total lepton charge asymmetry $l^{\pm}$ has a  different form when
expressed in terms of $r$ and $\bar{r}$ in the case of 
a coherently or incoherently produced $B\bar{B}$ pair.  When it is
produced coherently, for  example, in
$\Upsilon(4S)\rightarrow{B_d}\bar{B}_d$, the total lepton asymmetry is
given by \cite{franzini,hag,bss,yamamoto}
\begin{equation}
l^{\pm} ={l^+-l^- \over l^++l^-}={r-\bar{r} \over 2+r+\bar{r}}.
\end{equation}

One can  simplify 
the total lepton charge asymmetry $l^{\pm}$ by observing 
\begin{equation}
l^{\pm}=-\frac{|\eta|^4-1}
{\frac{4+2x^2}{x^2}|\eta|^2+|\eta|^4+1}\approx
0.17\ (A_{ll})_d\ {\rm for }\ B_d,
\end{equation}
here again we take $|\eta|\rightarrow{1}$ in the denominator.
In the $b$-factories, only $B_d\bar{B}_d$ is coherently produced.

When the $B\bar{B}$ is produced incoherently, the total lepton asymmetry 
becomes \cite{franzini}
\begin{equation}
l^{\pm}=\frac{r-\bar{r}}{1+r+\bar{r}+r\bar{r}}, 
\label{incoherent}
\end{equation}
which can also be simplified as 
\begin{equation}l^{\pm}=-\frac{|\eta|^4-1}
{(\frac{2+x^2}{x^2}+\frac{x^2}{2+x^2})|\eta|^2+|\eta|^4+1}\approx\left\{
\begin{array}{rl}
0.29\ (A_{ll})_d&{\rm for }\ B_d,\\
0.5\ (A_{ll})_s &{\rm for }\ B_s.
\end{array}\right.
\label{sindi}
\end{equation}

The single lepton asymmetry measures the same quantity, 
${\rm Im}(\Gamma_{12}/M_{12})$,
as the dilepton asymmetry. The prediction for the dilepton
asymmetry is bigger; however because both leptons must be tagged,
the statistics are smaller.
Yamamoto \cite{yamamoto} has argued that the single lepton
asymmetry will give a better measurement at a dedicated  $b$-factory.
At a hadron collider, the contribution from both $B_d$ and $B_s$
as well as other sources of single leptons must be accounted for:
\begin{equation}
(l^{\pm})_{\rm total}=l^{\pm}_{d}A_{d1}+l^{\pm}_{s}{A_{s1}},
\label{r2sum}
\end{equation}
where 
\begin{equation}
A_{d1}=\frac{\lambda_df_d}{\lambda^+f^++\lambda^{\Lambda}f^{\Lambda}+
\lambda_df_d+\lambda_sf_s},\hspace{0.2in}
A_{s1}=\frac{\lambda_sf_s}{\lambda^+f^++\lambda^{\Lambda}f^{\Lambda}+
\lambda_df_d+\lambda_sf_s},
\end{equation}
and $l^{\pm}_{d,s}$ takes the form of Eq.~(\ref{incoherent}).
If  the leptonic branching ratios were  the same 
for all  $b$-hadrons, $A_{d1,s1}$ would reduce to  $f_{d,s}$.
We define $A_1$ to the 
the ratio between $A_{s1}$ and $A_{d1}$, which is 
\begin{equation}
A_1=\frac{A_{s1}}{A_{d1}}=\frac{\lambda_sf_s}{\lambda_df_d}.
\end{equation}
For the values of $f_{d,s}$ and $\lambda_{d,s}$ given before, we get 
$A_{d1}=0.40$, $A_{s1}=0.11$, $A_1=0.26$.  However, to measure 
${\rm Im}\Gamma_{12}/M_{12}$ with
the same sensitivity for $B_d$ and $B_s$ requires the same factor
as with the dilepton asymmetry, in light of Eq.~(\ref{sindi}).

In the SM, the phases in $\Gamma_{12}$ and $M_{12}$ 
 are approximately equal.  Thus,
\begin{equation}
\Delta{M}_{\rm SM}\approx{2}|M_{12}^{\rm SM}|, \hspace{2cm}
\Delta\Gamma_{\rm SM}\approx{2}|\Gamma_{12}^{\rm SM}|.
\end{equation}
The SM contribution to the dilepton charge asymmetry is generally small
\cite{acuto,bss,xing1,nircp}
\begin{equation}
|A_{ll}|\sim\left\{
\begin{array}{cl}
{10}^{-3}&\mbox{for $B_{d}$},\\ 10^{-4}&\mbox{for $B_{s}$}.
\end{array}\right.
\end{equation}
The current preferred solutions of the unitary fits to the CKM matrix yield:
$\rho\simeq{0.12}$ and $\eta\simeq{0.34}$ \cite{ali}, 
which gives $|A_{ll}|\simeq{4.2}\times{10}^{-4}$
for $B_{d}$ in the SM.
When the mixing angles vary over their  
currently allowed  range \cite{nps}, 
the dilepton asymmetry can be as large as $1.9\times{10}^{-3}$ for $B_d$. 
Until the angles are
better known, a  lepton asymmetry in excess of this number is required
in order to  test new physics, so in this paper
this will be our benchmark. That is, we assume the
experimental sensitivity is good enough to measure
down to the largest possible standard model value.
However, once these angles are
better determined,   even  smaller  values could 
indicate non-standard
contributions.     We will show
that there are many interesting models
that predict a contribution at this level. 
The current experimental bound on the dilepton asymmetry is
$|A_{ll}|<0.18$ \cite{pdbold,cleo} at the 90\% confidence level, far
below the level of interest. We stress the
importance of  better measurements    at Run II of the Tevatron and at
$b$-factories,
with the ultimate goal of at least this sensitivity.

One should bear in mind the reduced contribution of $B_s$ relative
to $B_d$  (assuming equal ${\rm Im}(\Gamma_{12}/M_{12})$) which
means better experimental sensitivity is required to study
$B_s$ at the level assumed. This would require decoupling
any possible standard model contribution which could be present.

Another caveat is that the small standard model rate
is based on a quark calculation, and relies on a sensitive
cancellation between intermediate states with light up-type quarks.
Wolfenstein \cite{wolf} has argued that the quark model
calculations  might not be reliable and predicts a much larger
rate based on 100 \% violation of duality. Even with a smaller
violation of duality, of order 20 \% for the $c$ $\bar{c}$
intermediate state, we find the standard
model prediction could be increased by a factor of 3 if
the quark model rate is an overestimate. However, there
is no evidence as yet \cite {neubert} for this
violation. It will be interesting to better test the
assumption in the future by a better measurement
of  $n_c$, the average number of charm (or anticharm) quarks 
in the hadronic final state of a $B$ decay.

As is well known, there can be large FCNCs in supersymmetric models
because of the many new potentially flavor-violating  parameters. In
particular, the   squark mass matrices introduce the possibility of
new flavor-violating effects. These effects can be described through
the  mass matrices $\tilde{M}^2_{LL}$, $\tilde{M}^2_{RR}$, and
$\tilde{M}^2_{LR}$.  
Because of potential new contributions to  $B\bar{B}$
mixing, the phase of $\Gamma_{12}$ and 
$M_{12}$ should be different.  Assuming that  
supersymmetry does not substantially change 
$\Gamma_{12}$ (since it only contributes at higher order)
and defining $M_{12}^{\rm SUSY}/M_{12}^{\rm SM}
=h{\rm e}^{-{\rm i}\theta}$, the dilepton
asymmetry $A_{ll}$  is 
\begin{equation}
A_{ll}={\rm Im}\left(\frac{\Gamma_{12}^{\rm SM}}{M_{12}^{\rm
SM}+M_{12}^{\rm SUSY}}\right)=\left(\frac{\Delta\Gamma}
{\Delta{M}}\right)_{\rm
SM}\frac{h\sin\theta}{1+2h\cos\theta+h^2},
\label{r2}
\end{equation}
which is similar to a formula that was also presented in
Ref.~\cite{xing2}.
In  the SM, $(\Delta\Gamma/\Delta{M})_{\rm SM}$ is small 
\cite{hag,bss,bbd,cheng} :
\begin{equation}
\left(\frac{\Delta\Gamma}{\Delta{M}}\right)_{\rm
SM}=\left\{
\begin{array}{cl}
(1.3\pm{0.2})\times{10}^{-2}&\mbox{for $B_{d}$},\\
(5.6\pm{2.6})\times{10}^{-3}&\mbox{for $B_{s}$}.
\end{array}\right.
\label{sm}
\end{equation}
The errors break down as follows:  in $B_{d}$, $\pm{0.1}$ coming from
$m_{b}=4.8\pm{0.2}$ GeV; $\pm{0.1}$ coming from $m_{t}=165\pm{6}$ GeV;
$\pm{0.07}$ coming from the CKM matrix elements
$|V_{ud}V_{ub}^*|=0.003\pm{0.0008}$,
$|V_{cd}V_{cb}^*|=0.0086\pm{0.0007}$,
$|V_{td}V_{tb}^*|=0.0084\pm{0.0018}$ \cite{pdb}; 
$\pm{0.02}$ coming from $\eta_{\rm
QCD}=0.55\pm{0.01}$ \cite{buras}.  In $B_{s}$, the dominant error comes from
$B_{S}/B$. Here $B_{S}$ and $B$ are the ``bag'' parameters used to
estimate the matrix element 
$Q_S=(\bar{b}_i{s}_i)_{S-P}(\bar{b}_j{s}_j)_{S-P}$ and 
$Q=(\bar{b}_i{s}_i)_{V-A}(\bar{b}_j{s}_j)_{V-A}$ respectively 
(see Ref.~\cite{bbd}).  There are $\pm{2.3}$ from
$B_{S}/B$ varying between 0.7 and 1.3; $^{+1.1}_{-0.7}$ from varying
$\mu$ between $m_{b}/2$ and 2$m_{b}$; $\pm{0.4}$ from $m_{b}$ and
$\pm{0.4}$ from $m_{t}$.\footnote{The $m_{t}$ taken in Ref.~\cite{bbd}
is $m_{t}=176\pm{9}$ GeV.  It is slightly different from the $m_{t}$
we used in our calculation, which is the more recent experimental
result \cite{pdb}.}  The dilepton charge asymmetry  $A_{ll}$ can be
enhanced over the SM value by the second factor in
Eq.~(\ref{r2}). Notice that this factor reaches its maximum when $h=1$,
which gives
\begin{equation}
(A_{ll})_{\rm max}=\left(\frac{\Delta\Gamma}{\Delta{M}}\right)_{\rm SM}
\frac{1}{2}\tan\frac{\theta}{2}.
\label{rmax}
\end{equation}
This dilepton asymmetry is larger when $\theta$ is larger,
especially when $\theta$ is close to $\pi$.
\begin{figure}
\PSbox{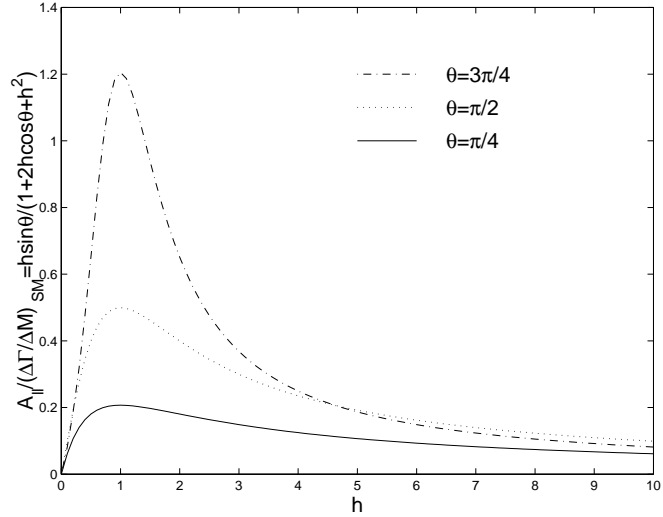 hoffset=72 voffset=-108 hscale=50  
vscale=50}{3.0in}{2.5in}
\caption{Plot  of dependence of  $A_{ll}/(\Delta\Gamma/\Delta{M})_{\rm SM}$ 
on $h$, with $\theta=\pi/4, \pi/2, 3\pi/4$  }
\label{r2h}
\end{figure}
We plot the dependence of $A_{ll}/(\Delta\Gamma/\Delta{M})_{\rm SM}$ on
$h$ by assuming the CP-violating phase difference $\theta$ to be $\pi/4,
\pi/2, 3\pi/4$ in Fig.~\ref{r2h}.  Notice that $A_{ll}$ is heavily
suppressed if $h$ is  either very large or very small. This is because
$\Delta M$ is large when $h$ is large,   while there is no significant
new contribution to $\Delta M$  when $h$ is small.  For a given
experimental sensitivity $(A_{ll})_{\rm min}$, there are corresponding
$h_{\rm min}$ and $h_{\rm max}$ (or none if $\theta$ is too small) to
which
the measurement is sensitive.  This  in turn gives
$(m_{\tilde{q}})_{\rm min}$ and $(m_{\tilde{q}})_{\rm max}$ by the
formulas
given in Sec.~\ref{numerical} for particular mixing angles.  A
measured dilepton asymmetry would constrain $m_{\tilde{q}}$ to be in a
range  between $(m_{\tilde{q}})_{\rm min}$ and 
$(m_{\tilde{q}})_{\rm max}$.   The
precise numerical results will be presented in Sec.~\ref{numerical}.

In the literature, there are two different parametrizations
of the effects of non-trivial squark matrices.
One can diagonalize the $\tilde{g}q\tilde{q}$
coupling and quark mass matrices while keeping all the mixing 
effects in the $\tilde{q}$
propagators.  This is called the ``mass insertion'' method \cite{hkr}.
One can also work in the mass eigenstates of quarks and squarks with 
off-diagonal gluino couplings (we call this the ``vertex mixing'' method), 
and consider only the contribution from the lightest generation. 
The ``mass insertion'' works well when the squarks are near degenerate, 
that is, for $\tilde{m}^2_{i}=\tilde{m}^2(1+x_{i})$, $i$=1--6 
for $\tilde{q}_L$ and $\tilde{q}_R$, $x_{i}\ll{1}$. 
Here $\tilde{m}$ is the average squark mass \cite{masiero}.
When the squark masses are not so degenerate but have the 
same order of magnitude, the mass insertion method can still be a good 
approximation if the average mass $\tilde{m}$ is chosen appropriately. 
The ``vertex mixing'' is a better approximation when one generation 
is much lighter than the other two since the contributions 
from the heavy generations 
are suppressed by their masses.
Notice that the mass insertion method assumes a GIM-like cancellation
of the leading term, which is why the results using vertex mixing
and mass insertion can be different.
We will use both methods in our
numerical calculation below, according to which is more appropriate.

In the flavor eigenstate basis 
of both quarks and squarks, the mass matrices for 
the up and down sector quark and squark are $M^{u}$, $M^{d}$,
$\tilde{M}^{u2}$ and $\tilde{M}^{d2}$, where the $6\times6$ 
squark mass matrix
$\tilde{M}^{d2}$ can be written in terms of $3\times3$ matrices
$\tilde{M}_{MN}^{d2}$ ($M,N=L,R$):
\begin{equation}
\tilde{M}^{d2}=\left(
\begin{array}{cc}
\tilde{M}_{LL}^{d2}&\tilde{M}_{LR}^{d2}\\ 
\tilde{M}_{RL}^{d2}&\tilde{M}_{RR}^{d2}
\end{array}\right).
\end{equation}
The off diagonal $\tilde{M}_{RL}^{d2}$ and $\tilde{M}_{LR}^{d2}$
are usually very small due to the suppression by
$m_{Z}/\tilde{m}$ and quark masses (in particular
for our purposes there is a $\lambda_b$ suppression).  In addition, 
the decay rate of $b\rightarrow{s\gamma}$ constrains 
$(\delta_{LR})_{23}$ (will be defined below) to be 
smaller than $1.6\times{10}^{-2}(m_{\tilde{q}}/500 {\rm GeV})^2$ 
\cite{masiero}. 
We neglect $\tilde{M}_{LR}^{d2}$ and diagonalize the squark
mass matrices $\tilde{M}_{LL}^{d2}$ and $\tilde{M}_{RR}^{d2}$  
in the mass basis of quarks and squarks, which defines the mixing angle $V$ by
\begin{eqnarray}
V_{L}^{d}M^{d}V_{R}^{d+}&=&{\rm diag}(m_d,m_s,m_b),\\
V_{L}^{u}M^{u}V_{R}^{u+}&=&{\rm diag}(m_u,m_c,m_t),\\
\tilde{V}_{L}^{d}\tilde{M}^{d2}_{LL}\tilde{V}_{L}^{d+}
&=&{\rm diag}(\tilde{m}_{dL}^2,\tilde{m}_{sL}^2,\tilde{m}_{bL}^2),\\
\tilde{V}_{R}^{d}\tilde{M}^{d2}_{RR}\tilde{V}_{R}^{d+}
&=&{\rm diag}(\tilde{m}_{dR}^2,\tilde{m}_{sR}^2,\tilde{m}_{bR}^2).
\end{eqnarray}
The $\tilde{g}q\tilde{q}$ vertices are in general
not  diagonal:   the coupling
mixing matrices (analogous to the standard CKM matrix) are
\begin{equation}
K_{L}^{d}=V_{L}^{d}\tilde{V}_{L}^{d+},\hspace{2cm}
K_{R}^{d}=V_{R}^{d}\tilde{V}_{R}^{d+},
\label{defk}
\end{equation}
and similarly 
for  the up system.  
This method of calculating flavor-changing effects is particularly
useful when the mass eigenstates are very non-degenerate.

We can also work in the basis where  the $\tilde{g}q\tilde{q}$ couplings
and
quark mass matrices are diagonal.  All the mixing  is now in the
squark propagators, which can be expressed in terms of the
dimensionless parameters $(\delta_{ij})_{MN}$
\begin{equation}
\delta_{MN}=\left(
\begin{array}{cc}
\delta_{LL}&\delta_{LR}\\ \delta_{RL}&\delta_{RR}
\end{array}\right)
=\frac{1}{\tilde{m}^2}\left(
\begin{array}{cc}
V_{L}^{d}\tilde{M}_{LL}^{d2}V_{L}^{d+}&V_{L}^{d}\tilde{M}_{LR}^{d2}V_{R}^{d+}\\
V_{R}^{d}\tilde{M}_{RL}^{d2}V_{L}^{d+}&V_{R}^{d}\tilde{M}_{RR}^{d2}V_{R}^{d+}
\end{array}\right).
\end{equation}
The mass insertion method is valid if  
$(\delta_{ij})_{MN}$ ($i\neq{j}$) is small.  

\section{ Models of Soft Supersymmetry Breaking}
\label{model}
\setcounter{equation}{0}
As is well known, the many additional parameters in supersymmetric
models can introduce large dangerous FCNC and CP-violating
effects. The parameters must be such that the
experimental 
bounds on $\varepsilon_{\rm k}$, $K$, $B$, $D$ mixing, the electric 
dipole moment (EDM) of electron ($d_{e}$) and 
the neutron ($d_{n}$) and branching ratio of $b\rightarrow{s}\gamma$ 
\cite{masiero,cplimit,hagelin} are preserved.  
Different scenarios have been proposed to solve, 
or at least relax the FCNC and CP-violation problems \cite{cpvio}.
It is possible that none of these
are the true solution, but they serve as useful straw men.
In this paper, we will discuss three of them: alignment
\cite{lns1,ns,nr,chmoroi}, non-abelian models 
\cite{chmoroi,bdh,bhrr,chmu1,chmu2,pt,dlk,ps,ks}
and heavy squark models \cite{dp,dg,ckn}. This list
of references is incomplete but incorporates
the models we study. Any model can be interpreted
as we do with these.  Models with nearly
exact universality, such as gauge-mediated models, are of course
an intriguing possibility for solving FCNC problems; however,
FCNC effects are generally  suppressed, and we therefore do not discuss
these here. In particular, the phase of $\Gamma_{12}$ and $M_{12}$
would still be correlated if universality were assumed
as a boundary condition.  Should the effects we describe be observed,
gauge-mediated models would be excluded.

The idea of alignment \cite{lns1,ns,nr} is that the squark mass matrices  
are aligned with the quark ones so that the $K_{L,R}$
in Eq.~(\ref{defk}) are close to the  identity; that is, the off-diagonal terms
are small.  Therefore, there are no large contributions to FCNC.  Such
models can be constructed with an abelian horizontal symmetry $H$
and additional scalar fields $S$ \cite{lns2}.  With the appropriate
assignment of the horizontal quantum numbers to $S$, 
Higgs fields $\phi_{u,d}$ and matter
fields $Q, d, u$, one can construct non-renormalizable terms in the Lagrangian
\begin{equation}
\frac{\lambda_{ij}^d}{M^{m_{ij}}}Q_{i}\phi_{d}S^{m_{ij}}\bar{d}_{j}
+\frac{\lambda_{ij}^u}{M^{n_{ij}}}Q_{i}\phi_{u}
S^{n_{ij}}\bar{u}_{j}+{\rm h.c.},
\end{equation}
which can give masses to the fermions if $S$ has a vacuum expectation
value $\langle{S}\rangle$.  Here $m_{ij}$ and $n_{ij}$ are determined
by the $H$ charge assignment, so that the terms are invariant under
$H$, $M$ is a higher energy scale that communicates the horizontal
symmetry breaking to the light states, and  $\lambda_{ij}^{d,u}$ are
some coefficients of order 1.   The horizontal symmetry  is
spontaneously broken when $S$ gets a vacuum expectation value, which
introduces a small number $\epsilon=\langle{S}\rangle/M$ (this is the
Froggatt-Nielson mechanism \cite{fraggott}).  Different powers of
$\epsilon$ in the Yukawa coupling  account for  mixing  angles and
the hierarchy of  fermion masses.  In supersymmetric theories, squarks
have the same $H$ charges as the quarks of the same multiplet and will
obtain masses by the same mechanism.  An astute choice of charges can
allow for  the alignment of squark mass matrices  with  quark mass
matrices, thereby suppressing flavor-changing effects.  Notice that
more than one U(1) symmetry is  generally  needed in order to get
feasible models consistent with the experimental bounds.  In \cite{nr},
the alignment model is associated with spontaneous  CP-violation,
which can predict small values of supersymmetric  CP-violating phases
so that the EDM bounds are satisfied.  After diagonizing both  the
quark and squark mass matrices, we found that the (12) and (13) mixing
angles in the gluino coupling vertices (although the (12) mixing angle 
is too small for this to be relevant) and the (13)  component of the
CKM matrix   can have a CP-violating phase $O(1)$,  
while the (23) mixing is almost
real,  so there will be  no sizable contribution to the dilepton
charge asymmetry for $B_{s}$.  

The non-abelian models
\cite{chmoroi,bdh,bhrr,chmu1,chmu2,pt,dlk,ps,ks}  are motivated by the
large top mass and the different behavior of the third family with 
respect to 
the first two families.   The maximum flavor symmetry group is
${\rm U}(3)^5$ in the absence of Yukawa couplings.  It can be assumed that
there is a non-abelian flavor symmetry $G_f \subset {\rm U}(3)^5$ where the
first two families and the third family transform differently;   
$G_f$ can be continuous or discrete, gauged or
global.  There are a variety of models based on different $G_{f}$:
$G_{f}={\rm U}(2)$ \cite{bdh,bhrr,pt}, $G_{f}={\rm SU}(2)$ \cite{chmoroi,dlk},
$G_{f}=({\rm S}_3)^3$ \cite{chmu1,chmu2}, 
$G_{f}={\rm U}(1)\times{{\rm O}(2)}/{\rm Z}_{2}$
\cite{ps} and $G_{f}=\Delta(75)$ \cite{ks}.  This symmetry is only
approximate, and  is broken by some small factor $\lambda$.  Because
the symmetry guarantees that the 
first two families are    nearly degenerate,  
FCNC for the light generations are heavily suppressed.  
In some models where the
degeneracy of the first two families does not fully resolve the FCNC
constraints \cite{bdh,pt,dlk,ps}, different scenarios  are proposed to
relax the constraints:  the first two generations can be much heavier
than the third generation $\tilde{m}_{1,2}\sim{10}\tilde{m}_{3}$ 
(scenario (a) in 
\cite{pt}), 
or the CP-violation phase is very small
$\varphi\sim{10}^{-2}$ (scenario (b) in \cite{pt} and \cite{ps}).

The heavy squark models \cite{dp,dg,ckn} provide another possible
solution to the FCNC and CP-violation problems by allowing
the maximal masses consistent with naturalness bounds.  All those
models permit the first two generations of squarks to be heavier,
which is crucial for solving  FCNC problems.   In
effective supersymmetry \cite{ckn},  a new gauge group $G$ is
introduced, which enlarges the accidental symmetry group and thus
forbids the renormalizable $B$- and $L$-violating interactions.  It
also introduces a new mass scale $\tilde{M}\sim$ 5--20 TeV, which sets
the mass scale for the first two generations.  In addition, the
requirement of naturalness implies that some squarks ($\tilde t_{\rm
L}, \tilde{b}_{\rm L}$) and most gauginos must have a mass  below
$\sim 1$  TeV.  In \cite{dp}, the first two families are charged under 
a gauged ${\rm U}(1)$ symmetry, while the third family is neutral.  
Thus, there is an extra contribution to squark masses of the the first two 
families coming from the 
D-term, which generates the mass hierarchy.
The constraints on the mixing
angles come  from the naturalness of the Higgs sector and the
squark mass matrices \cite{ckln}.  We should emphasize that the heavy-squark 
models cannot fully satisfy the FCNC and CP constraints by
themselves.  They have to be combined with non-abelian symmetry
(scenario (a) in \cite{pt}) or have some alignment  for the squark 
mass matrices.

In the next section, we select  specific models from the 
above-mentioned papers 
and  present a  detailed study of the dilepton asymmetry
and  how it can put constraints on the squark masses,
mixing angles and CP-violation phases for these models. Of 
particular interest are
those models that specified the squark and quark mass matrix textures 
(therefore we do not consider \cite{ks})
and have  large mixing with the third generation (so we do not consider
\cite{chmu2}). 
We also require that the models we consider  satisfy the FCNC and 
CP-violation bounds set by the experimental value of  $\varepsilon_{K}$,
$\Delta{M}_{K}$, $\Delta{M}_{D}$, $d_{e}$ and $d_{n}$.  This excludes
models in \cite{bdh,dlk}, and  bounds  the squark masses in
some other models.  In \cite{nr}, $m_{\tilde{q}}>200$ GeV, so that the
quark-squark alignment solution to $\Delta{m_{K}}$ will not run into
problems with  $\Delta{m_{D}}$.  In the scenario (a) of \cite{pt},
$m_{Q_{3}, D_{3}}$ should be heavier than 550 GeV if
$m_{\tilde{q}}\sim{m_{\tilde{g}}}$, so that the supersymmetric 
contribution to
$\varepsilon_{K}$ is within the experimental bounds, but this constraint
is relaxed if the CP-violation phase is small, so we do not impose this bound.
In \cite{ckn},
there is 1 TeV upper bound which comes from the 
naturalness of the  Higgs sector.
For the models where the squarks of the first two generations are heavy
(scenario (a) in \cite{pt} and \cite{ckn}), 
$\tilde{m}_{1,2}$ are required to be heavier than $10\tilde{m}_{3}$.

The models should have an $O(1)$ CP-violation phase difference in 
the (13) and (23) mixing angles between the  SM and supersymmetric models,
so that their contributions to the dilepton
asymmetry is not negligible.  Thus, in \cite{nr}, we only consider (13) mixing
since the (23) mixing CP-violation phase difference is small.

\begin{table}
\begin{tabular}{clccccccc} \hline
&&\multicolumn{3}{c}{(23) mixing}&\multicolumn{3}{c}{(13)
mixing}\\\cline{3-8}   &Model&$LL$&$RR$&&$LL$&$RR$&&\\ \hline
&\cite{lns1}&$\lambda^2$&$\lambda^4$&$LL\gg{RR}$&$\lambda^{3}$&
$\lambda^{3}$&$LL=RR$&\\
A&\cite{ns}, \cite{chmoroi} a&$\lambda^2$&1&$LL\ll{RR}$&\multicolumn{3}{c}{Too
small mixing}&\\ &\cite{nr} &\multicolumn{3}{c}{Small CP-violation
angle}&$\lambda^{3}$& $\lambda^{7}$&$LL\gg{RR}$&Mass\\\cline{1-8}
&\cite{chmoroi} b&$\lambda^2$&$\lambda^{1/2}$&$LL\ll{RR}$&
$\lambda^{3}$&$\lambda^{3/2}$&$LL\ll{RR}$&insertion\\ B&
\cite{bhrr}, \cite{pt} b
&$\lambda^2$&$\lambda^2$&$LL=RR$&$\lambda^{3}$&
$\lambda^{3}$&$LL=RR$&\\
&\cite{chmu1}&$\lambda^3$&$\lambda^5$&$LL\gg{RR}$&$\lambda^{2}$&
$\lambda^{4}$&$LL\gg{RR}$&\\
&\cite{ps}&$\lambda^2$&$\lambda^4$&$LL\gg{RR}$&$\lambda^3$&$\lambda^4$&
$LL\sim{RR}$&\\
\hline B$+$C&\cite{pt} a
&$\lambda^2$&$\lambda^2$&$LL=RR$&$\lambda^{3}$&
$\lambda^{3}$&$LL=RR$&Vertex\\ \cline{1-8}
C&\cite{ckn}&$\lambda^2$&&$LL\gg{RR}$&$\lambda^3$&
&$LL\gg{RR}$&mixing\\\hline
\end{tabular}
\caption{Selected models from the literature, which will 
be analysed in section~\ref{numerical}.  
Here A --- alignment, B --- non-abelian, C --- heavy squarks.}
\label{modeltable}
\end{table}
In Table \ref{modeltable}, we list  different models  with the predicted
$LL$ and $RR$ mixing (up to $O(1)$ uncertainties).  As 
will be shown in the next section, $M_{12}$ can get
contributions from both $LL$ and $RR$ mixing.  There are two different
cases:
\begin{enumerate}
\item Either $LL$ or $RR$ mixing dominates.
\item $LL$ mixing is comparable with $RR$ mixing.
For definiteness, we take $LL=RR$ in
our calculation, except for model in ref.~\cite{ps}, where the (13) 
$LL$ and $RR$ mixing are $\lambda^3$ and $\lambda^4$ respectively, which is 
denoted by $LL\sim{RR}$ in Table~\ref{modeltable}.  It is nonetheless important
to take the $RR$ term into account  
because the contribution to $M_{12}$ from 
$(\delta_{31})_{LL}(\delta_{31})_{RR}$ 
is large.
\end{enumerate} 
We also show in the table whether the mass insertion (with mixing
parameter $\delta_{LL,RR}$) or the vertex mixing (with mixing
parameter $K_{L,R}$) method is used.  We only use the vertex
mixing in scenario (a) of \cite{pt} and in \cite{ckn}, 
where the first two generations are much
heavier than the third one, while the mass insertion should be  a good
approximation in  the other cases, though it should be noted
that in a detailed analysis one would account for the nondegeneracy
of the squarks.

\section{Numerical Results}
\label{numerical}
\setcounter{equation}{0} 
We will assume the 
measurement of ${\rm Im}(\Gamma_{12}/M_{12})$ for 
either $B_d$ or $B_s$ with a sensitivity $2\times{10}^{-3}$.
This measurement can be obtained from either the single 
or dilepton asymmetries. 
For definiteness, since $A_{ll}\sim{\rm Im}(\Gamma_{12}/M_{12})$, we 
refer to the dilepton asymmetry in this section.

From Eq.~(\ref{r2}), we see that the dilepton
asymmetry depends on $(\Delta\Gamma/\Delta{M})_{\rm SM}$,  $h$ and
$\theta$.  $(\Delta\Gamma/\Delta{M})_{\rm SM}$ is given in
Eq.~(\ref{sm}) which has been calculated in the SM;
$\theta$ is the phase difference between $M_{12}^{\rm SM}$ and 
$M_{12}^{\rm SUSY}$,
which can be in the range from  0 to $2\pi$. Since $|A_{ll}|$ is symmetric
with respect to $\pi$,  we only consider $\theta$ to be in the range
of 0 to $\pi$. The quantity
 $h$ is the ratio between the amplitudes of $M_{12}^{\rm
SUSY}$ and $M_{12}^{\rm SM}$, which can be calculated through the
$\Delta{B}=2$ box diagrams with $\tilde{q}, \tilde{g}$ or $q, W$
running in the loop\footnote{Here we do not consider the contribution
from box diagrams with $\tilde{q}$ and chargino (or neutralino)
running in the loop since it is  suppressed by a factor
$(g_2/g_s)^4$ with respect to the $\tilde{q}, \tilde{g}$  
box diagram contribution.};
$M_{12}^{\rm SM}$ has already been calculated, including QCD
corrections \cite{franzini,bss,cheng}, and gives (we only include the
top-quark contribution since this is the largest effects):
\begin{equation}
M_{12}^{\rm SM}=\frac{G_{F}^2}{12\pi^2}m_t^2B_{B_{q}}f_{B_{q}}^2
M_{B_{q}}(V_{tq}V_{tb}^*)^2\frac{A(z_t)}{z_t}\eta_{\rm QCD},
\label{smm}
\end{equation}
where 
\begin{equation}
\frac{A(z_t)}{z_t}=\frac{1}{4}+\frac{9}{4}\frac{1}{1-z_t}-
\frac{3}{2}\frac{1}{(1-z_t)^2}-\frac{3}{2}\frac{z_t^2{\rm ln}z_t}
{(1-z_t)^3}, \hspace{0.5cm}z_t=\frac{m_t^2}{m_W^2}.
\end{equation}
$B_{B_{q}}$ is the ''bag'' parameter describing the uncertainty in
evaluation of the hadronic matrix element, $M_{B_{q}}$ and
$f_{B_{q}}$ are the $B_{q}$ meson mass and decay constant
respectively.  Although $B_{B_{q}}f_{B_{q}}^2M_{B_{q}}$
cancels in $h$, as will be shown below, 
we still need to know the value of $B_{B_d}$ and $f_{B_d}$ 
since we have to take into account the constraint from
$\Delta{M}_{B_d}$.
The recent values of $B_{B_d}$ and $f_{B_d}$ are 
$B_{B_d}=1.29\pm{0.08}\pm{0.06}$ \cite{gimenez},
$f_{B_d}=175\pm{25}$ MeV \cite{flynn}.  $\eta_{\rm QCD}$ is the QCD
correction factor, which is taken  to be $0.55\pm0.01$ \cite{buras} in
our calculation.   $M_{12}^{\rm SUSY}$ can be calculated either in the
scenario of vertex mixing \cite{hagelin} or using mass insertion
\cite{masiero}.  In the first case, $M_{12}^{\rm SUSY}$ is given in
terms of $K_{L,R}^d$ (we take $(K_{L,R}^d)_{33}\sim{1}$)\footnote{
There are mistakes in the formulas given by Ref.~\cite{hagelin}, which is
pointed out by Ref.~\cite{masiero}.  We use the corrected formulas here.}:
\begin{eqnarray}
\lefteqn{M_{12}^{\rm
VM}=-\frac{\alpha_s^2}{216m_{\tilde{q}}^2}
\frac{1}{3}B_{B_{q}}f_{B_{q}}^2M_{B_{q}}
\left\{((K_{L}^d)_{3i}^2+(K_{R}^d)_{3i}^2)
(66\tilde{f}_4(x)+24xf_4(x))+\right.}\nonumber
\\
&&\hspace*{-0.4cm}\left.(K_{L}^d)_{3i}(K_{R}^d)_{3i}
\left[\left(36-24\left(\frac{M_{B_{q}}}{m_b+m_q}\right)^2\right)
\tilde{f}_4(x)+\left(72+384\left(\frac{M_{B_{q}}}{m_b+m_q}\right)^2
\right)xf_{4}(x)\right]\right\},\nonumber\\
\label{cohen}\\
&&f_4(x)=\frac{2-2x+(1+x){\rm ln}x}{(x-1)^3},\hspace{0.5cm}
\tilde{f}_4(x)=\frac{1-x^2+2x{\rm ln}x}{(x-1)^3},
\hspace{0.5cm}x=m_{\tilde{g}}^2/m_{\tilde{q}}^2,
\end{eqnarray}
where $i=1,2$ for $B_d$ and $B_s$ respectively.
In the mass insertion notation,  
\begin{eqnarray}
\lefteqn{M_{12}^{\rm
MI}=-\frac{\alpha_s^2}{216m_{\tilde{q}}^2}
\frac{1}{3}B_{B_{q}}f_{B_{q}}^2M_{B_{q}}
\left\{((\delta_{3i})_{LL}^2+(\delta_{3i})_{RR}^2)
(66\tilde{f}_6(x)+24xf_6(x))+\right.}\nonumber
\\
&&\hspace*{-0.2cm}\left.(\delta_{3i})_{LL}(\delta_{3i})_{RR}
\left[\left(36-24\left(\frac{M_{B_{q}}}{m_b+m_q}\right)^2\right)
\tilde{f}_6(x)+\left(72+384\left(\frac{M_{B_{q}}}{m_b+m_q}\right)^2
\right)xf_{6}(x)\right]\right\},\nonumber\\
\label{ma}\\
&&\hspace*{-0.2cm}f_6(x)=\frac{6(1+3x){\rm ln}x+x^3-9x^2-9x+17}{6(x-1)^5},
\\
&&\hspace*{-0.2cm}
\tilde{f}_6(x)=\frac{6x(1+x){\rm ln}x-x^3-9x^2+9x+1}{3(x-1)^5}.
\end{eqnarray}
The mass parameters we used in our calculation below is
$M_{B_d}=5279.2\pm{1.8}$ MeV, $M_{B_s}=5369.3\pm{2.0}$ MeV,
$m_{W}=80.41\pm{0.10}$ GeV \cite{pdb}.  From Eqs.~(\ref{smm}),
(\ref{cohen}) and (\ref{ma}), we can see that $h$ depends only  on
$m_{\tilde{q}}$, mixing parameters, and $x$.  The $x$ dependence
comes from the functions $f_4(x)$, $\tilde{f}_4(x)$ or $f_6(x)$,
$\tilde{f}_6(x)$.
\begin{figure}
\PSbox{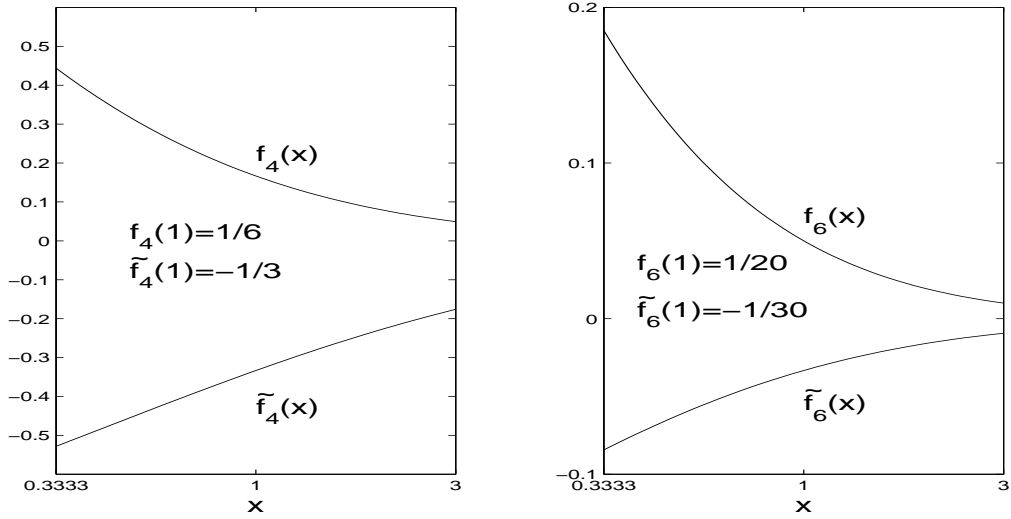 hoffset=0 voffset=-108 hscale=80
vscale=50}{6.0in}{2.5in}
\caption{Plot of  $f_4(x)$, 
$\tilde{f}_4(x)$, $f_6(x)$, 
$\tilde{f}_6(x)$with $x$ in the range of 1/3--3. }
\label{ftotal}
\end{figure}
We plot the $x$ dependence of these functions in Fig.~\ref{ftotal} 
and also show their values when $x=1$, which is the value of
$x$ we use  in the calculation below.  The supersymmetric contribution
measured by $h$  can dominate  over the SM prediction for
$\Delta{M}_{B_{q}}$ because of large mixing angles or small
$m_{\tilde{q}}$. In Eq.~(\ref{r2}), we see that $A_{ll}$ is suppressed
when $h$ is  either too large or too small; thus  the limit on the
value of $A_{ll}$ that will be experimentally accessible translates
into a range of $h$ that can be probed, given a specific
$\theta$. This in turn determines the range of $m_{\tilde{q}}$ that
can be probed in particular models, which specify (approximately) the
mixing angles.  Again we emphasize that in general for detectable
values of $\theta$ we will find both an upper and lower bound on $h$
which will be experimentally accessible. The upper bound corresponds
to too large mixing whereas the lower bound corresponds to too small a
supersymmetric contribution (so too small CP violation). The upper
limit on $h$ translates into a lower bound on the squark mass (for
definite mixing angles) whereas the lower bound translates into an
upper bound on the mass.

For our results we take into account
the constraint
$\Delta{M}_{B_{d}}=0.470\pm0.019\ {\rm ps}^{-1}$ \cite{pdb}.
The supersymmetric contribution to $\Delta{M}_{B_{d}}$ 
cannot exceed this, 
which puts a severe bound on the (13) mixing and $m_{\tilde{q}}$.
The $B_s$ mixing has no such constraint.

As explained in Sec.~\ref{formula}, $B_{d}$ and $B_{s}$ both contribute
to the total dilepton rate.
Assuming a measurement on the dilepton asymmetry with a sensitivity  
of $10^{-3}$, we require at least
$2\times{10}^{-3}$ dilepton asymmetry from either $B_{d}$ or $B_{s}$.
This is also the current upper bound on the $B_d$ standard model
contribution, so it is clearly identifiable as new physics.
Of course if better precision is possible a measurement of the $B_s$
system with greater accuracy would be interesting since its
standard model rate is much lower.

In Eq.~(\ref{rmax}), we see that the phase difference 
between $M_{12}^{\rm SUSY}$ and $M_{12}^{\rm SM}$
must be large enough so that the dilepton asymmetry 
can be measured.  For the precision
we choose, we find that 
\begin{equation}
\theta\geq\left\{
\begin{array}{cc}
30^{\circ}-40^{\circ}&{\rm for}\ B_{d},\\
52^{\circ}-106^{\circ}&{\rm for}\ B_{s}.
\end{array}\right.
\end{equation}
Here the large range in the lower limit of 
$\theta$ comes from the large uncertainties in 
$(\Delta{\Gamma}/\Delta{M})_{\rm SM}$.
Models with too small a CP-violation phase difference between the SM and 
supersymmetric models  cannot be 
tested using the dilepton asymmetry. 

For the results presented below, we take
the SM CKM matrix elements \cite{pdb,hein}:
\begin{equation}
|V_{td}V_{tb}^*|:\hspace{0.5cm}0.0066-0.0102,\hspace{2cm}
|V_{ts}V_{tb}^*|:\hspace{0.5cm}0.026-0.060.
\end{equation}

\begin{table}
\begin{tabular}{ccccc} \hline
&\multicolumn{2}{c}{Vertex mixing}&
\multicolumn{2}{c}{Mass insertion}\\\hline
$\theta$&$LL$ (${RR}$) dominates&
$LL\sim{RR}$&$LL$ (${RR}$) dominates&$LL\sim{RR}$\\ \hline
$\pi/2$&$m_{\tilde{q}}\sim${192--505}&
$m_{\tilde{q}}\sim${323--849}
&$m_{\tilde{q}}\sim${45--119}&$m_{\tilde{q}}\sim${224--590}\\
$3\pi/4$&$m_{\tilde{q}}\sim${192--485}&
$m_{\tilde{q}}\sim${323--814}
&$m_{\tilde{q}}\sim${45--114}&
$m_{\tilde{q}}\sim${224--566}\\ 
\hline
\end{tabular}
\caption{$m_{\tilde{q}}$ (GeV) range from $B_{d}$ dilepton 
asymmetry with  
$A_{ll}=2\times10^{-3}$, $x=m_{\tilde{g}}^2/m_{\tilde{q}}^2=1$ and the 
mixing angles are the same as the corresponding CKM entry. }
\label{msqlimit1}
\end{table}

\begin{table}
\begin{tabular}{ccccc} \hline
&\multicolumn{2}{c}{Vertex mixing}&
\multicolumn{2}{c}{Mass insertion}\\\hline
$\theta$&$LL$ (${RR}$) dominates&$LL\sim{RR}$&
$LL$ (${RR}$) dominates&$LL\sim{RR}$\\ \hline
$\pi/2$&$m_{\tilde{q}}\sim${130--310}&$m_{\tilde{q}}\sim${214--509}
&$m_{\tilde{q}}\sim${31--73}&
$m_{\tilde{q}}\sim${150--357}\\
$3\pi/4$&$m_{\tilde{q}}\sim${115--352}&
$m_{\tilde{q}}\sim${188--578}
&$m_{\tilde{q}}\sim${27--83}&
$m_{\tilde{q}}\sim${132--405}\\ 
\hline
\end{tabular}
\caption{$m_{\tilde{q}}$ (GeV) range from $B_{s}$ dilepton 
asymmetry with  
$A_{ll}=2\times10^{-3}$, $x=m_{\tilde{g}}^2/m_{\tilde{q}}^2=1$ 
and the mixing angles are the same as the corresponding CKM entry. }
\label{msqlimit2}
\end{table}

In Tables \ref{msqlimit1} and \ref{msqlimit2}, we show the allowed ranges
of $m_{\tilde{q}}$ when the dilepton asymmetry of  $B_{d}$ or
$B_{s}$ 
is larger than $2\times{10^{-3}}$.  Here  we
assume that the (13), (23) mixing angles are the same as the
corresponding CKM entries.  
We notice that the vertex mixing results are larger than
the mass insertion ones.  This is because we only consider the
contribution from the third generation in the vertex mixing case,
while all the generations contribute in the mass insertion
method.  In the latter case, the leading-order contribution  cancels
because of the GIM mechanism, which gives smaller $M_{12}^{\rm SUSY}$.
Thus, the squark masses obtained using the mass insertion method
have to be smaller in order to compensate for  this weakening effect.
However without a detailed knowledge of the mixing angles and masses
either method must be viewed as an approximation.

We need to take into account the
current experimental lower bounds on the squark mass coming  from the
non-observation of any supersymmetry signals at either LEP
\cite{grivaz} or the Tevatron \cite{tev1,d0}.  For the vertex mixing
case, when the first two generation squarks are much heavier than
the third generation ones,  the $m_{\tilde{q}}$ bounds are on the
lightest sbottom mass. The sbottom masses are  all larger than 110
GeV, which have not been excluded by the experimental lower limit
($m_{Z}/2$) \cite{grivaz}.   For the mass insertion case when
the squark masses are almost  degenerate, we impose the constraints
from CDF and D\O\ searches.  While $LL$ (or $RR$) dominates, the
squark masses  are smaller than 120 GeV, which  has already been ruled
out by the CDF ($m_{\tilde{q}}>230$ GeV \cite{tev1}) and the D\O\
($m_{\tilde{q}}>260$ GeV\cite{d0}) limit; this means that if the soft
supersymmetry-breaking models have the squark mixing angles in the
SM ranges and $LL$ (or $RR$) is dominating, then the
dilepton asymmetry is too small to be measured experimentally.  If
$LL=RR$, the squark masses can be in the range of 130--600 GeV;
most of the mass ranges have not been ruled out yet.  Thus, the
corresponding supersymmetric models can still be tested through a
dilepton charge asymmetry measurement.

Also notice that the lower limits of $m_{\tilde{q}}$ in
Table~\ref{msqlimit1} ($B_{d}$ case) are the same for different
$\theta$.  This is because the upper bound on $h$  (which corresponds to
the lower bound on $m_{\tilde{q}}$) is set by the experimentally
measured value of $\Delta{M}_{B_{d}}$.

\begin{table}
\begin{tabular}{clcccccccc} \hline
&&\multicolumn{4}{c}{(23) mixing}&\multicolumn{4}{c}{(13)
mixing}\\\cline{3-10}   &&&&\multicolumn{2}{c}{$m_{\tilde{q}}$
(GeV)}&&&\multicolumn{2}{c}{$m_{\tilde{q}}$ (GeV)}\\
\cline{5-6}\cline{9-10}
&Model&$LL$&$RR$&$\theta=\pi/2$&$\theta=3\pi/4$&$LL$&$RR$&
$\theta=\pi/2$&$\theta=3\pi/4$\\ \hline
&\cite{lns1}&$\lambda^2$&$\lambda^4$&20--112&18--127&$\lambda^{3}$&
$\lambda^{3}$&176--715&176--686\\
A&\cite{ns}, \cite{chmoroi} a&$\lambda^2$&1&511--2807&450--3187&
\multicolumn{4}{c}{Too
small mixing}\\ &\cite{nr} &\multicolumn{4}{c}{Small CP-violation
angle}&$\lambda^{3}$& $\lambda^{7}$&36--144&36--139\\\cline{1-10}
&\cite{chmoroi} b&$\lambda^2$&$\lambda^{1/2}$&229--1255&201--1425&
$\lambda^{3}$&$\lambda^{3/2}$&397--1614&397--1549\\ B&\cite{bhrr}, \cite{pt} b
&$\lambda^2$&$\lambda^2$&100--549&88--624&$\lambda^{3}$&
$\lambda^{3}$&176--715&175--686\\
&\cite{chmu1}&$\lambda^3$&$\lambda^5$&4--22&4--25&$\lambda^{2}$&
$\lambda^{4}$&178--722&178--693\\
&\cite{ps}&$\lambda^2$&$\lambda^4$&20--112&18--127&$\lambda^3$&
$\lambda^4$&73--298&73--286
\\\hline
B$+$C&\cite{pt} a
&$\lambda^2$&$\lambda^2$&143--784&126--890&$\lambda^{3}$&
$\lambda^{3}$&253--1029&253--987\\ \cline{1-10}
C&\cite{ckn}&$\lambda^2$&&87--476&76--541&$\lambda^3$&
&151--613&151--588\\\hline
\end{tabular}
\caption{$m_{\tilde{q}}$ ranges for different models with 
$A_{ll}=2\times10^{-3}$, $x=m_{\tilde{g}}^2/m_{\tilde{q}}^2=1$  and
$\lambda=0.2$ for $B_{s}$ ((23) mixing) and $B_{d}$ ((13) mixing)
dilepton asymmetries. Notations are the same as
in Table~\ref{modeltable}.}
\label{modelresults}
\end{table}

We now consider the implications of our results.
In Table~\ref{modelresults}, we show the ranges of $m_{\tilde{q}}$ for
different soft supersymmetry-breaking models.  
The lower mass limits coming from the constraints of FCNC have been imposed.   
We see that
the models divide into several groups: those that show an
effect in both, neither, or one of the $B_d$ and $B_s$ systems.
It should of course be remembered that we used a definite value
for the angle in these tables so there is an order one fudge
factor available in any model. Nonetheless, there are some clear tendencies
indicated by these results. 

For the  models in \cite{bhrr,pt}, both the $(\delta_{13,23})_{LL}$ and 
$(\delta_{13,23})_{RR}$ are of 
the order of $\lambda^3$ and $\lambda^2$, respectively, which are
comparable to the SM CKM mixing angles.  This
is the optimal situation in that for reasonable supersymmetric
masses, the standard and nonstandard box contributions
are competitive.  The squark masses
are below 1 TeV, which can be explored  experimentally.

The models in \cite{ns,chmoroi} predict quite different mixing
angles from the SM.  The $RR$ mixing can be as large as 1
in (23) and $O(\lambda^{3/2})$ \cite{chmoroi} in (13), which
increases the squarks to heavy masses in order to keep the 
supersymmetric and SM box diagrams to be comparable to 
each other ($h\sim{1}$).
If the squark
masses are indeed light, it is likely that there is a large
supersymmetric contribution to flavor violation and that it
will be completely undetected in the measurement of a lepton
asymmetry! Since such large mixing angles will give rise
to noticeable effects in other measurements, the lepton
asymmetry provides a nice complement to such measurements
and would provide a clear determination of a model with
large mixing angles. However, it is possible that the squark
masses are at the lighter end (which would be measured)
and that there is visible evidence of a lepton asymmetry.

There are several models which should give a reasonable
asymmetry for $B_d$ but not for $B_s$. These include the models
in \cite{lns1,chmu1,ps}, although the latter only has a small
range of squark mass which has not already been excluded by the CDF
and D\O\ limits. Actually for \cite{nr}, the
squark mass range which is accessible
is already at the limit of what would
be permitted once the 200 GeV lower
bound arising from the 
$\Delta{m}_{D}$ constraint is accounted for.
We can see that the model of Ref. \cite{nr} which
explicitly addresses the CP violation as well as the flavor structure
gives no measurable effects in the lepton asymmetries.

We also find that the heavy squark models are very likely testable.
However, we note that we assumed the mixing angles agreed with their
CKM counterparts. Should they be bigger, the asymmetry could be
too small to measure.  For the
effective supersymmetry model in \cite{ckn}, 
the upper limit increases with the mixing
angles; it is 600 GeV for the mixing to be in the SM
ranges.  If the mixing is too large, this model predicts 
too small dilepton asymmetry because of the 1 TeV upper limit  in squark
masses coming from the naturalness constraints and thus cannot be 
tested.

In the analysis above, we assumed that the dilepton asymmetry of
$2\times{10}^{-3}$ is measured for either $B_{d}$ or $B_s$.  If we
assume $4\times{10}^{-3}$ sensitivity, then the lower limit on
$\theta$ is  $56^{\circ}-72^{\circ}\ {\rm for}\ B_{d}$ and
$89^{\circ}-139^{\circ}\ {\rm for}\ B_{s}$.  
For any given experimental sensitivity $D$, we define a 
$\theta$-dependent scale factor
$S_{d,s}(\theta,\frac{D}{2\times{10}^{-3}})$.
The squark mass under this new sensitivity is then
$(m_{\tilde{q}})_{D}=S_{d,s}(\theta,\frac{D}{2\times{10}^{-3}})
(m_{\tilde{q}})_{2\times{10}^{-3}}$.
If we take $\theta=3\pi/4$ and $D=4\times{10}^{-3}$ 
as an example, in $B_d$, the lower
limit remains the same because it comes from the $\Delta{M}_{B_d}$
constraint, while for the upper limit $S_d(3\pi/4,2)_{\rm upper}=0.765$.  
In $B_s$,
the reduction in squark mass range follows from  
$S_s(3\pi/4,2)_{\rm upper}=0.78,\ S_s(3\pi/4,2)_{\rm down}=1.28$.
In general, the factor $D$ can be obtained from Eq.~(\ref{r2}).

We conclude that models are certainly distinguishable
from the lepton asymmetry measurements alone. With other
complementary measures there is some hope to resolve the flavor
problem of supersymmetry. However, we have assumed a reasonably
good sensitivity, which is essential for the measurement to be useful.

\section{Conclusions}
\label{conclusion}
It is possible that the dilepton asymmetry could
be one of the first indications of physics beyond the standard model.
Once the source of new physics is ascertained through direct measurements, 
it can be used to impose further
constraints. In particular, if the new physics is indeed
low-energy supersymmetry, the 
dilepton charge asymmetry can be used to distinguish various soft
supersymmetry-breaking models. The range of parameters which
can be tested is such that there is often a good overlap with
interesting flavor models. Unfortunately, there is not an
unambiguous identification of the size of the signal with
the category of model; nonetheless particular models
with definite patterns for masses and mixings can be tested.
We emphasize the importance of taking a model-dependent approach;
although large mixing angles can give big effects in searches
for physics beyond the standard model, existing constraints
and the attempt to motivate flavor physics parameters
through an underlying symmetry structure often disfavor this assumption.

We emphasize that there are uncertainties in the precise
angles which can change the exact range of squark mass which is
covered. In particular, all models
we presented have order unity uncertainties
for the angles which can change the
range of mass which is probed.
Furthermore nondegeneracy of the squark mass and gluino
mass introduces another parameter which can affect the precise
range of parameters which is covered. Finally,
the mass insertion method is an approximation; it
is generally only  a good one if the mass
parameter is appropriately interpreted.  Nonetheless, the 
overall message is clear; it would be very interesting to
do an accurate measurement of the single lepton and/or dilepton
asymmetries.

In this paper, we have focussed on the lepton charge asymmetry as a
means of searching for new physics, in particular for testing new models
of flavor. These lepton asymmetries are sensitive to new sources of
mixing in the $B$ system.  These new sources of mixing can be
independently tested in other measurements
\cite{nirslac,dib,ckln,newmeasure,worah}. Particularly interesting
alternative tests for CP-violation in $B_s$ are the study of $B_s \to
D_s^+D_s^-$, $B_s \to \psi \phi$, and related modes
\cite{nircp,bs,bds}; $B_d$, in  particular modes such as $B_d \to
\phi K_s$  and $B_d \to K_s \pi^0$ \cite{worah,bds,bd}, where new
physics effects might be large, and measurements of deviations in
precisely predicted rates such as $B \to J/\psi K_s$.  In particular,
models with large mixing in the $B_s$ sector (of order unity) should
give large effects for these measurments.
Even with these studies, it will
be useful to have an alternative means of searching for new
physics. Precisely because the rate is negligible in the standard
model, the lepton asymmetry would be an extremely important measurement.

\noindent{\bf Acknowledgements:} We are grateful to U. Aglietti,
M. Lusignoli, G. Martinelli,   A. Nelson, and especially
Y. Nir, R. Rattazzi,  D. Wood,
and H. Yamamoto  for useful conversations.

\end{document}